# Moss-like growth of metal electrodes: On the role of competing faradaic reactions and fast-charging

J.X. Kent Zheng[1,2], Jiefu Yin[3], Tian Tang[1], Lynden A. Archer[1,3*]

1. Department of Materials Science and Engineering, Cornell University, Ithaca, NY, 14853, USA.

2. Department of Physics, Massachusetts Institute of Technology, Cambridge, MA, 02129, USA.

3. Robert Frederick Smith School of Chemical and Biomolecular Engineering, Cornell University, Ithaca, NY, 14853, USA.

\* Corresponding author: laa25@cornell.edu

**Abstract**: Uncontrolled crystal growth during electroreduction of reactive metals in liquid electrolytes produces porous, low-density, mossy metal deposits that grow primarily along the surface normal vector to a planar electrode substrate. The moss-like deposits are fragile and cause premature failure of batteries by chemical, physical, and mechanical pathways. Here we use electroanalytical Rotating-Disk Electrode (RDE) studies in a three-electrode electrochemical cell to elucidate the fundamental origin of moss-like growth of metals. We report that competing Faradaic reactions occurring on the electrode surface is the source of the phenomenon. On this basis, we conclude that a moss-like growth regime can be accessed during electrodeposition of any metal by subtle shifts in electrolyte chemistry and deposition rate. Specifically, for Zn—a metal that conventionally is not known to form moss-like electrodeposits—obvious moss-like deposition patterns emerge at low-current densities in strongly-alkaline electrolytes that undergo electroreduction to form an interphase on the electrodeposited Zn. Conversely, we find that under conditions where the rate of metal electroplating is large relative to that of other competing Faradaic reactions, it is possible to eliminate the mossy-like growth regime for Zn. Motivated by these observations, we evaluate the principle in Li metal batteries and again find that the typical moss-like electrodeposit morphology of Li can be eliminated. Li metal deposits in this regime are uniform and particulate, leading to very high Li plating/stripping reversibility (>99.9%) over 1500 cycles. We also study the practical utility of our findings in Li||LiFePO$_4$ full cells (areal capacity ≈ 3 mAh/cm$^2$) and show that lifetimes of at least 350 cycles are achieved at high current densities of 70 mA/cm$^2$; meaning that the cells can be fully charged in < 5 minutes. Taken together, our findings open up a new, surprising approach for simultaneously achieving favorable metal deposition morphology and fast charging in next-generation batteries using metal anodes.

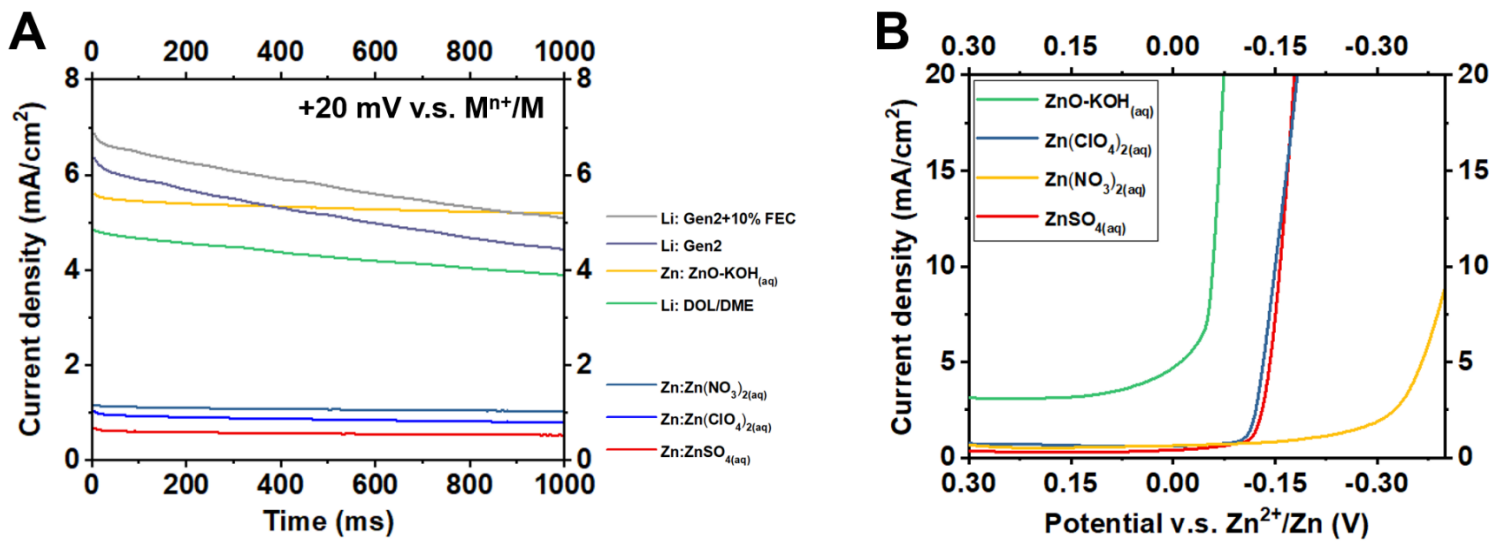

**Figure 1. Electrochemical evaluation of the Faradaic reactions occurring at Li and Zn electrodes in different electrolytes.** (A) Chronoamperometry measurements of Li and Zn in various liquid electrolyte systems. The potential is held at +20 mV versus the redox potential of the metal electrode (*i.e.*, $Zn^{2+}/Zn$ or $Li^+/Li$). (B) Linear sweep voltammetry measurements of typical Zn electrolyte systems. Scan rate: 20 mV/s.

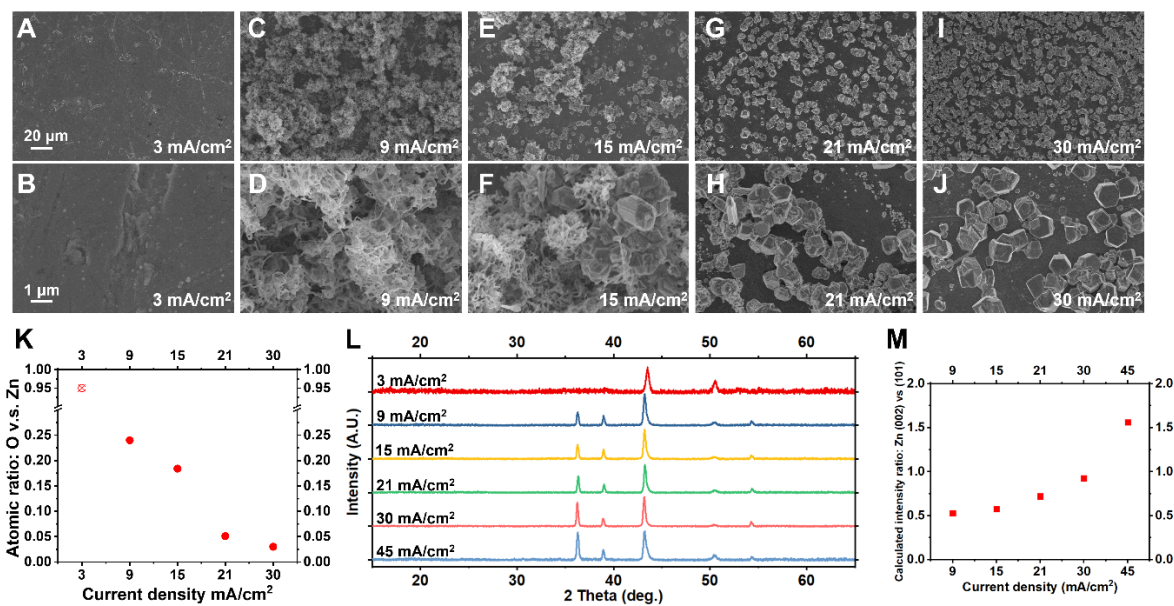

**Figure 2. Microstructural and chemical characterization of Zn electrodeposits grown in a strong SEI-forming, alkaline electrolyte (30 wt% KOH saturated with ZnO).** (A)~(J) Zn electrodeposition morphology obtained at different current densities, as specified at the lower right corner of the images. See more SEM images of the moss-like Zn electrodeposits in **Fig. S3**. (K) EDS analysis of Zn and O of the electrodeposits. (L) XRD patterns of the electrodeposits. (M) Estimated ratio between $(002)_{Zn}$ intensity and $(101)_{Zn}$ intensity of the electrodeposits. The diffraction generated by stainless steel at around $2\theta \approx 43°$ was subtracted when estimating the $(002)_{Zn}:(101)_{Zn}$ ratio.

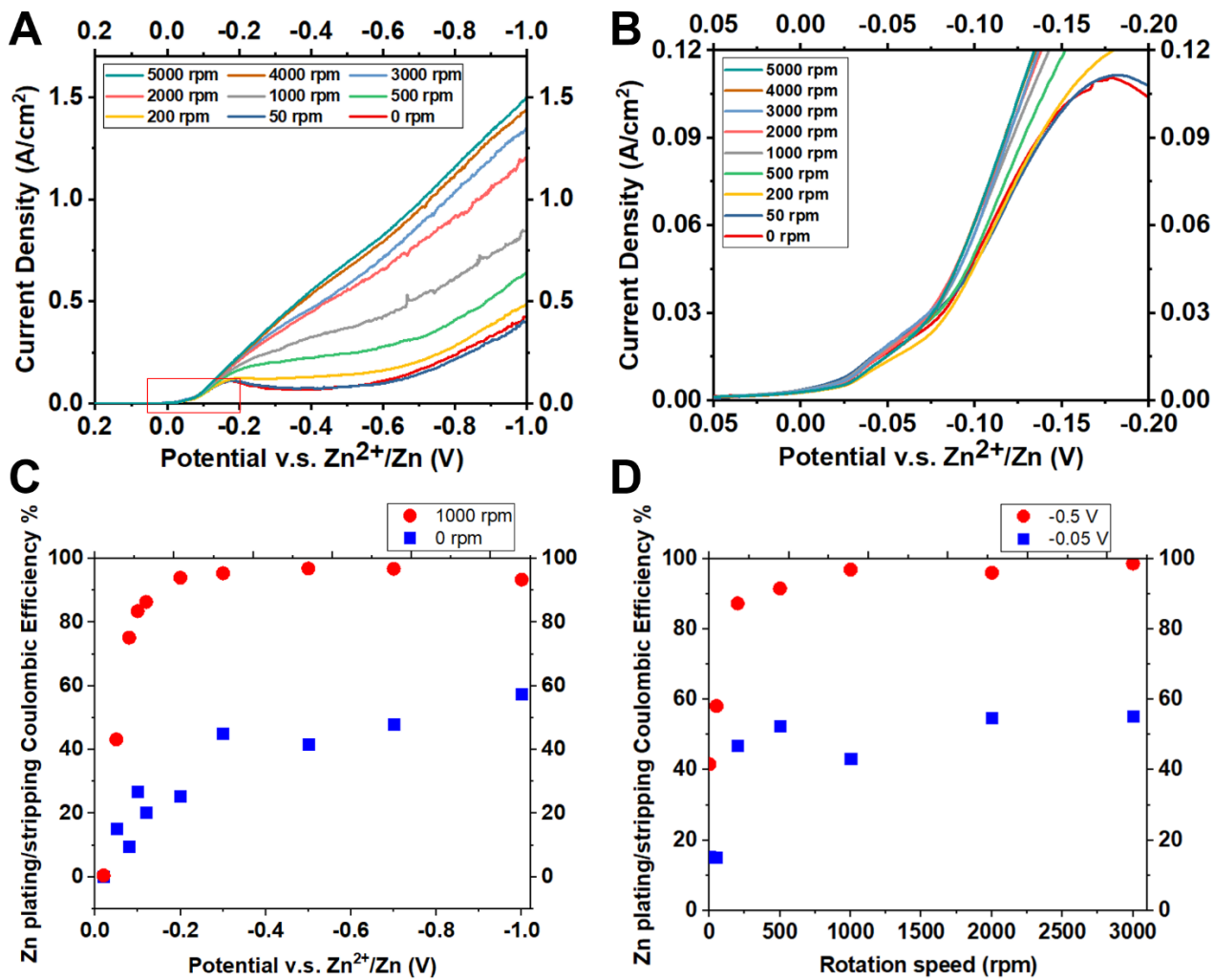

**Figure 3. Zn plating/stripping in the strongly SEI-forming electrolyte (30 wt% KOH saturated with ZnO) at a rotating disk electrode.** (A) Linear sweep voltammetry measurements under different electrode rotation speeds. (B) Magnified plot showing the i-V plot near the onset of Zn plating. Dependence of Zn plating/stripping Coulombic efficiency on potential at fixed rotation speeds (C), and at a fixed overpotential and variable electrode rotation speeds (D).

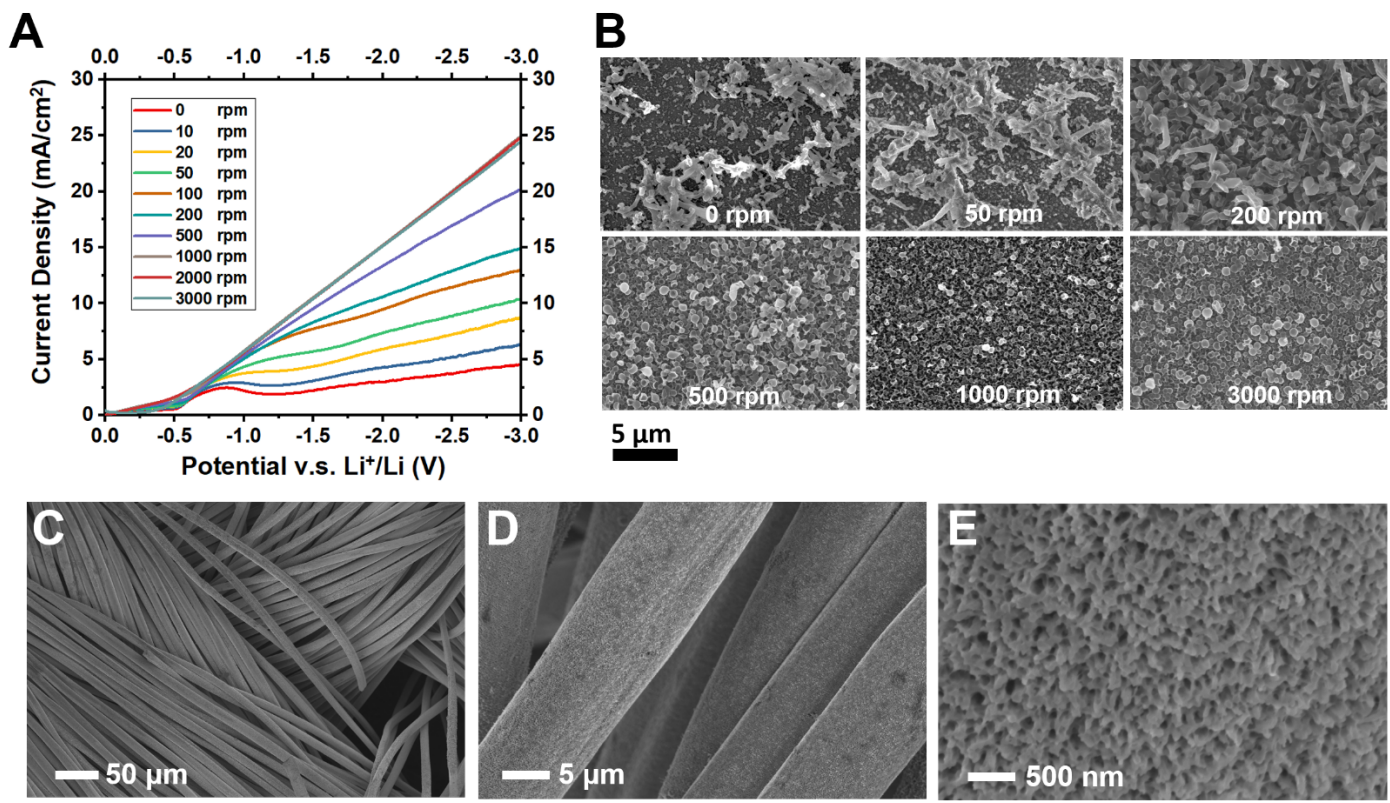

**Figure 4. Controlling the growth morphology of Li electrodes.** (A) Linear sweep voltammetry measurements of Li plating at a rotating disc electrode, under different rotation speeds. Electrolyte: 50 mM LiPF$_6$ in EC/DMC with 10% FEC. (B) SEM images of Li metal deposition morphology at -2.0 V versus Li$^+$/Li at a rotating disc electrode at different rotation speeds. (C)(E) Electrodeposition morphology of Li on a nonplanar substrate at -1.6 V versus Li+/Li (areal capacity: 0.5 mAh/cm$^2$; current density: ~10$^2$ mA/cm$^2$) in a stagnant electrolyte. Electrolyte: 1 M LiPF$_6$ in EC/DMC with 10% FEC.

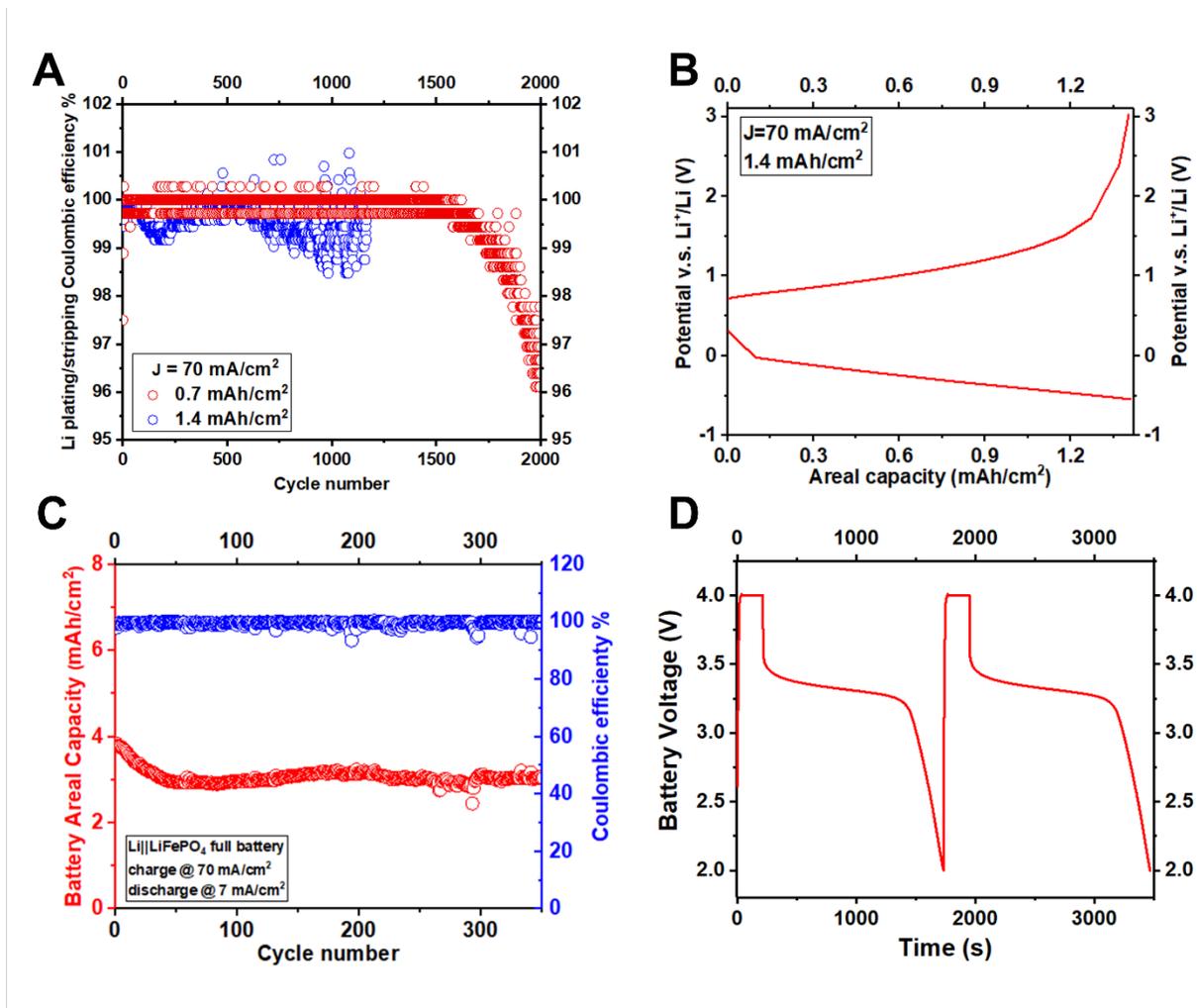

**Figure 5. Electrochemical reversibility of Li metal anodes at high rates.** (A) Li plating/stripping Coulombic efficiency in a 1 M LiTFSI DOL/DME electrolyte obtained at 70 mA/cm$^2$, and (B) the corresponding voltage profile. **Notice the large expansion of the y-axis scale required to see the differences beyond 1500 cycles**. (C) Li||LiFePO$_4$ full battery cycling performance using a charging current density of ~70 mA/cm$^2$, and (D) the voltage profiles of the 100$^{th}$ and 101$^{st}$ cycles. The cell was charged to and then held at 4 V until the current density drops

from 70 mA/cm$^2$ to 30 mA/cm$^2$. Electrolyte: 1 M LiPF$_6$ in EC/DMC with 10% FEC. A carbon-fiber cloth is used as anode current collector in all these cells.

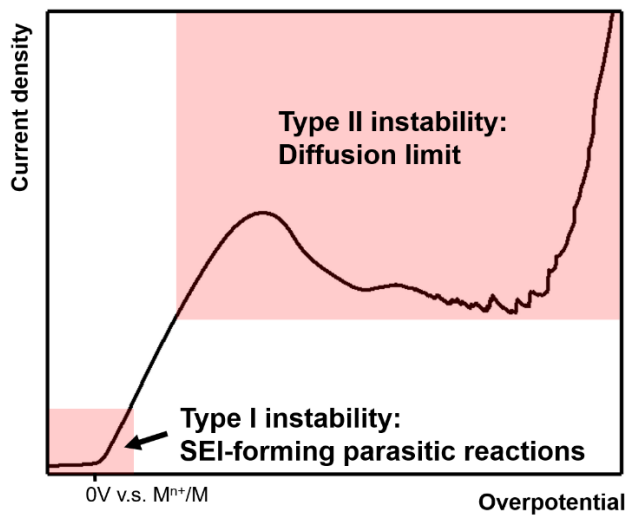

**Scheme 1. Illustration showing the two instabilities that drive heterogeneous growth of metals in battery anodes.** Type I and Type II instabilities manifests themselves as moss-like and dendrite-like growth modes, respectively.

**Main text begins:**

"**D**endritic" growth of metal electrodeposits at planar substrates is now a well- known failure mode for secondary batteries and in electroplating processes. The term *dendritic* deposition has also been used in contemporary literature to loosely refer to a variety of fundamentally different electrodepositon processes that produce metal deposits with preferred growth normal to the plane and typically with porous, mossy morphologies.[1,2] Dendrites, by etymology, are tree-like/fractal structures that are the norm for crystal growth in systems that are both near or driven far from equilibrium by external fields. For example, tree-like growth of a solid phase characterized by self-similarity is unambiguously understood in a large number of classical contexts to arise from transport limitations or symmetry, *e.g.*, alloy casting,[3] ice forming,[4] and traditional electroplating.[5] A large body of theoretical and experimental work has shown that in dilute electrolytes, electroplating of most metals at current densities above the diffusion limit for the metal cations produces classical dendritic deposits.[6] On this basis, one would expect the dendritic growth to be eliminated at low current densities below the diffusion limit. This hypothesis has been tested extensively for Li metal deposition in battery anodes from a variety of perspectives, and the results are at best confusing—Li metal consistently deposits into porous, non-planar structures composed of tortuous nano/micro-wires even at a very small current density, and across vastly different electrolyte chemistries.[7] And, more detailed microscopy studies reveal that the porous, non-planar structures do not follow a tree-like growth mode, in contrast to dendrites observed in classical metal electroplating processes. These structures are then best described as "*moss*-like" Li as opposed to *dendritic*. This distinction is of course fundamental because it means that whereas dendritic deposition of a metal can be prevented by a variety of process and design variables (*e.g.*, by maintaining the current density below certain limits; by adding a supporting electrolyte salt with a cation that is electrochemically inactive at the potentials at which metal plating or stripping occur; by using such large salt concentrations that the double layer thickness in the electrolyte is comparable to the thickness of the cation depletion zone at the diffusion limit, *etc*.), no such general strategies exist for ameliorating moss-like deposition.

In the context of next-generation batteries using Li metal anodes, the benefits of understanding mossy deposition are well-known and discussed at length in a rising number of incisive literature reviews.[8,9] Briefly, the moss-like growth is now known to be detrimental to battery operation for a number of reasons: it causes fast capacity fading, propensity for internal battery short, consumption of electrolyte and so on. While the research of Li metal anodes over the past 10 years has revealed a number of materials design strategies that overcome these limitations in one way or another and have resulted in steady improvements in the plating/stripping reversibility of Li anodes from impractical low values to reliably above 90%, achieving the 10% in battery cell and electrolyte designs that do not compromise the beneficial aspects of Li is essential for progress towards all practical post-Li ion batteries. Understanding and eliminating the moss-like Li growth is critical to achieving this goal.[10] Answering this question could also generate profound implications in fundamental sciences. The inability to understand and predict the shape of crystals grown under specified conditions represents a major gap in knowledge. We note further that the moss-like growth pattern observed during electroreduction of metals has also been reported in materials synthesis using totally different methods, *e.g.*, vapor deposition.[11]

The large volume of prior literature focused on improving Li metal anodes provide a good foundation for investigating the origins of mossy metal deposition. Failure to stop the process — despite the range of methods proposed, attests to the robustness of the underling processes and may even mean that the moss-like growth regime is actually the preferred growth mode under typical conditions.[7] A second set of insights come from the fact that a mossy deposition regime exists for all other alkali metals like Na and K and in essentially all electrolytes. In contrast, mossy electrodeposits are normally not observed for non-alkali metals, *e.g.,* Al, Mg, Zn, Cu, etc. electrodeposited in neutral electrolytes. The moss-like growth then evidently originates from a differentiating property of the alkali metal systems.

There are at least three categories of hypotheses advanced in the literature to explain why alkali metals form dominantly non-planar electrodeposits: (a) those based on surface diffusivity, (b) those based on cation transport rate in the electrolyte bulk, (c) those based chemical instability/reactivity of the electrolyte.[9] The issue is unsettled, which has led to a proliferation of literature approaching the issue of moss-like growth from different angles, *e.g.,* using a substrate that has optimal Li surface diffusivity,[12, 13] choosing an electrolyte that forms a more uniform interphase on Li,[14] employing a three-dimensional host to reduce the current density,[7] respectively. While these strategies are capable of improving the overall battery performance, their influence on the microstructure of the Li electrodeposits remains unresolved.[7]

A complementary question is whether ***non***-alkali metals that usually do not form moss-like deposits, *e.g.*, Zn, can be driven into a similar moss-like regime as alkali metals? We show later that identifying the special set of conditions that allow any metal to transition to moss-like growth offers critical insights about both the origin of the moss-like growth of alkali metals and how it might be eliminated. Zn also stands out as a good system for in-depth studies complementary to those performed in alkali metals for a number of reasons: as already noted, Zn does not form mossy deposits in neutral aqueous electrolytes; Zn deposition in strongly alkaline aqueous electrolytes is known to be challenged by interphase formation and large interfacial impedances; as a hexagonal crystalline material, Zn has distinctive crystallographic growth characteristics in comparison to alkali metals, which form cubic crystal lattices; finally Zn deposition can be carried out in aqueous media where inorganic compounds are the dominant and typically only interphase components formed by electroreduction of electrolyte components, this contrasts with the more complex organic-inorganic hybrid interphases formed on alkali metal anodes in aprotic organic electrolytes.

**Results**

The analyses above suggest that experimental efforts focused on a 2D plane—with deposition rate and electrolyte chemistry being the two dimensions, could provide a fruitful approach for identifying the dominant mossy growth regime(s) in any metal. The deposition rate is easily controllable by setting the applied current. We first focus on the effect of electrolyte chemistry. The propensity of a given electrolyte chemistry for undergoing parasitic reactions can be evaluated using analytical electrochemical characterization tools. **Fig. 1A** reports the current response of typical Li and Zn electrolytes in chronoamperometric measurements. The potential of the working electrode is held at +20 mV versus the redox potential of the metal ($Li^+/Li$ or $Zn^{2+}/Zn$). It should be noted that the presence of an overpotential is one requirement for metal electrodeposition, which corresponds to negative potentials on the working electrode. At this slightly positive potential of +20 mV, there is no driving force for metal electrodeposition; the measured current should instead be attributable to parasitic side reactions. The magnitude of the observed response current therefore serves as a both a qualitative and quantitative indicator of the propensity of an electrolyte for decomposition at electrochemical potentials near the redox potential of the metal.

A rather large response current >5 mA/cm$^2$ is observed for all three Li metal systems studied. The response current densities are in fact comparable to typical current densities used in contemporary studies of galvanostatic Li electrodeposition, *i.e.*, <2 mA/cm$^2$. Under these conditions, the measured voltages of Li cells reportedly remain above 0 V versus $Li^+/Li$ at the initial stage.[15] The observation of the significant response current at +20 mV versus $Li^+/Li$ provide a straightforward explanation for the above-zero voltage capacity — it also belies the strong propensity of the electrolytes to undergo parasitic reactions and to continuously form SEI at voltages close to the onset potential of Li electrodeposition.

We next consider the behaviors of various Zn electrolytes at slightly positive potentials (relative to $Zn^{2+}/Zn$ ). The alkaline Zn electrolyte[16] composed of 30 wt% KOH saturated by ZnO shows a response current density comparable in magnitude to those observed for Li, but nearly one order of magnitude higher than observed for the neutral aqueous electrolytes studied. The large current density observed in the alkaline electrolyte is attributable to parasitic reactions. We note that this same alkaline electrolyte has been widely used in Zn battery studies. Of particular interest is that—similar to the Li electrolytes—the parasitic reactions are believed to generate solid products (*e.g.*, ZnO)[17, 18] that precipitate onto and cover the electrode surface. Based on the high impedance of a ZnO interphase, we speculate that Zn electrodeposits from the alkaline electrolytes may have a higher probability than other Zn electrolytes to develop a moss-like morphology, analogous to the Li systems. And the alkaline Zn electrolyte offers a unique, powerful platform for interrogating the role played by chemical instability and SEI formation in the electrodeposition of Zn.

To further investigate the Faradaic processes occurring in the regime close to the onset of Zn plating/stripping, we performed linear sweep voltammetry (LSV) measurements in the Zn electrolytes (**Fig. 1B**). Consistent with the chronoamperometry tests, the alkaline Zn electrolyte exhibits the highest parasitic reaction current in the positive potential range (i.e., from 0.30 to 0 V versus $Zn^{2+}/Zn$). Another significant finding is that the faradaic current above 0 V versus $Zn^{2+}/Zn$ originating from the parasitic reactions is not sensitive to the overpotential. It in fact remains around 3 mA/cm$^2$ over the full positive potential range. This insensitivity contrasts starkly with the nearly exponential current-voltage relationship observed after the onset potential (slightly negative) for Zn electroreduction is reached. We believe that this difference is central to understanding the competing nature of the faradaic reactions in electrodeposition, and how this competition influences the electrodeposit growth morphology. The ratio between the deposition current and the parasitic reactions current can be manipulated by imposing different overpotentials in a potentiostatic electrodeposition or current densities in a galvanostatic deposition. For example, at a smaller overpotential/current, the parasitic reactions are more prominent, resulting in a faster rate of SEI formation relative to metal deposition. The balance is reversed at high overpotentials/currents.

On this basis, one would expect that moss-like growth is most likely at low current/overpotential regime and can be arrested in the high current/overpotential regimes. We note that this expectation is exactly opposite to the conclusion one would reach using any of the transport-based hypotheses—a high current/overpotential will give rise to moss/wire-like growth. **Figure 2A~J** reports the morphologies of Zn electrodeposits obtained in the alkaline electrolytes at different current densities (i.e., 3, 9, 15, 21, 30 mA/cm$^2$, respectively). As evidenced by these SEM images, the growth pattern of Zn is sensitive to the current density. At 3 mA/cm$^2$, the potential is above 0 V versus Zn$^{2+}$/Zn (**Fig. S1**), meaning that the Faradaic current observed should not be attributed to Zn deposition. Instead, parasitic reactions contribute the majority—if not the whole—of the measured current. In particular, no reversible plating/stripping behavior is detected in Coulombic efficiency measurement under the low current density condition (**Fig. S2**). This is also confirmed in the SEM characterization—no Zn deposit is observable. Instead, a number of oxygen-enriched small particles are detected; and EDS analysis shows the O:Zn atomic ratio is close to 1:1, suggesting that ZnO is the dominant interphase species, which is consistent with prior literature.[17,]

[18] This means that either no Zn is formed or the Zn deposits are so porous/loose that they detach easily from the substrates during the deposition process. Taken together, these results suggests that—surprisingly— the galvanostatic electrochemical response of Zn in alkaline aqueous electrolytes is dominated by chemical instability at low current densities.

Interestingly, the Zn deposits show a characteristic moss-like morphology at a higher current density 9 mA/cm$^2$; see high-magnification images in **Fig. S3**. As the current density increases, the Zn morphology transitions gradually to a more well-defined crystalline morphology composed of particulate crystallites. We note that the shapes of the particles follow those deduced from the theoretically predicted Wulff construction for the hexagonal close packed Zn crystal. Specifically, at 15 mA/cm$^2$, the growth pattern is a mixture of the moss-like deposits and the crystalline deposits. See also optical microscopy and 2D height profiling in **Fig. S4~S5**. Whereas at 30 mA/cm$^2$, the crystalline particles dominate! This morphological evolution of Zn over the applied current density is in good agreement with earlier studies.[19, 20] Of note is that in tandem with the transition to crystalline Zn electrodeposits, the O:Zn ratio continues to decrease from 0.24:1 at 9 mA/cm$^2$ to 0.03:1 at 30 mA/cm$^2$ (**Fig. 2K**). These results support the conclusion that the SEI has a more pronounced effect at lower currents than at higher currents (*i.e.*, content and effect of SEI: 9 mA/cm$^2$ > 15 mA/cm$^2$ > 21 mA/cm$^2$ > 30 mA/cm$^2$).

The morphological transition seen in **Figs. 2A-2J** is also reflected in part by the crystallographic features of the electrodeposits. Consistent with the microstructure characterization, at 3 mA/cm$^2$, only the diffraction peak of the stainless-steel substrate is detectable, suggesting parasitic reactions dominate the Faradaic process under this condition. Starting from 9 mA/cm$^2$, characteristic diffraction peaks corresponding to $(002)_{Zn}$, $(100)_{Zn}$ and $(101)_{Zn}$ are detected. A key finding is that the $(002)_{Zn}$ diffraction intensifies as the current density increases; this point can be clearly seen from the plot in **Fig. 2M**, which shows the ratio between $(002)_{Zn}$ and $(101)_{Zn}$. This ratio in fact serves as a rough estimate of the out-of-plane texture of the deposits. The randomness associated with crystallographic texture of the deposits formed at low current densities is a manifestation of the random orientation and alignment of the individual Zn wires in the moss-like electrodeposits. The intensification of the $(002)_{Zn}$ at a relatively higher current density is consistent with what is reported in less SEI-forming electrolytes, *e.g.*, $ZnSO_{4(aq)}$, where Zn crystallites tend to align the basal (002) plane horizontally with respect to the electrode surface.[21] These results about the Zn morphology developed in the alkaline electrolyte should be compared to the surface morphology formed in other Zn electrolytes that exhibit smaller Faradaic contributions from parasitic reactions (**Fig. 1** and **Fig. S6**).

Next we used a rotating disk electrode (RDE) to resolve the chemical and the transport aspects involved in the deposition process (**Fig. 3**). The rationale for using a RDE is manifold. The RDE facilitates detailed electrochemical analysis in a three-electrode configuration, which facilitates more precise conclusions than possible from the two-electrode systems, e.g., coin cells. Additionally, electrochemical measurements in an RDE allows the mass transport near the electrode to be precisely controlled by an artificial convective flow: as expressed by the Levich equation,[22, 23] the approximate thickness of the hydrodynamic boundary layer can be calculated as $\delta = \frac{D_0^{1/3} v^{1/6}}{0.620 \omega^{1/2}}$, where $D_0$ is the diffusivity, v is the kinematic viscosity of the solution and ω is the angular rotation rate. **Figs. 3A & 3B** summarize the main LSV measurements for Zn in alkaline electrolytes. At rotation speeds ranging from 0 to 5000 rpm the system exhibits a typical i-V response to the negative potential sweep. As shown in the magnified plot in **Fig. 3B**, the response starts with a regime controlled by electrochemical kinetics, captured by the Butler-Volmer equation, before entering the diffusion-limited regime with a limiting current density around 100

mA/cm$^2$. Of particular note is that the i-V relation in the initial regime shows negligible dependence on the rotation speed. It means that mass transport does not play an important role in the electrochemical process within this range (i.e., <100 mA/cm$^2$). We examined the Zn electrodeposit morphology on the working electrode of the RDE (**Fig. S7**). The results are evidently consistent with those obtained from the coin-cell experiments reported in **Fig. 2**. Taken together, these observations show that the morphological transitions of Zn electrodeposits at low current densities (*e.g.*, in **Fig. 2**) cannot be attributed to mass transport related origins.

It is believed that the plating/stripping reversibility of metal electrodeposits are fundamentally correlated to their morphology. Motivated by this hypothesis, we measured the plating/stripping Coulombic efficiency of the Zn electrodeposits formed under different overpotential and rotation conditions (**Fig. 3C** and **3D**). At low overpotentials where Zn forms moss-like deposits, the plating/stripping Coulombic efficiency is consistently low (*i.e.*, <90%), regardless of the rotation conditions. As the overpotential increases, the system however quickly enters the diffusion limited regime, which results in porous, chemotactic growth pattern and incomplete stripping [21]. In this regime, rotation-induced convective flow provides effective enhancement in mass transport and remarkably improves the plating/stripping efficiency. These findings based on Zn metal (**Fig. 1~3**) unveil the two types of instabilities of fundamentally different origins as illustrated in **Scheme 1**: the excessive SEI formation due to parasitic reactions at the low current/overpotential regime, and the diffusion limit at the high current/overpotential regime, respectively. Stable electrodeposition is achievable at moderately high currents/overpotentials away from both instabilities, or at high currents/overpotentials with enhanced mass transport—for example—enabled by artificial convective flow.

We next apply the insights gained from studying Zn to understand the origin of moss-like Li growth. The current densities used for Li deposition in contemporary literature is < 5mA/cm$^2$, sometimes <1 mA/cm$^2$. Comparing these values to current densities associated with the parasitic reactions reported in **Fig. 1A**, it is reasonable to conclude that the parasitic reactions that form heterogeneous SEI play a key role in the formation of moss-like Li within the low current/overpotential regime. Approaches for circumventing the moss-like Li growth can therefore be devised leveraging these understandings. **Figure 4A** reports the LSV investigation of Li deposition from a carbonate-based electrolyte at different rotation speeds. Analogous to the Zn deposition from the alkaline electrolyte, the i-V response of this Li electrolyte transitions from a first regime (at low currents/overpotentials) where it is relatively insensitive to rotation speed. This is followed by a second regime at larger currents/overpotentials where the i-V response is highly sensitive to the rotation speed. Based on the understandings gained from the Zn case, we hypothesize that simultaneously applying a large overpotential and imposing a rotation-induced convective flow would provide a mechanism for obtaining smooth-planar Li deposits in any electrolyte. **Fig. 4B** summaries the Li electrodeposit morphology obtained at a large overpotential of -2 V versus Li$^+$/Li at different rotation speeds. Without rotation, the Li metal undergoes an outward, chemotaxis-type growth due to mass transport limit as previously reported.[21] As the rotation speed increases, this outward growth is gradually suppressed, and completely eliminated at rotation speeds greater than 1000 rpm. Surprisingly, under this condition, the Li metal forms uniform deposition layer composed of sphere-shape particles! This is in stark contrast to the moss-like morphology reported consistently in prior literature [7] (see also **Fig. S8**). The successful elimination of the moss-like growth pattern of Li using the method of high potential plus rotation further supports that both of the instabilities illustrated in **Scheme 1** must be simultaneously suppressed in order to promote uniform, compact deposition morphology desirable in rechargeable battery anodes.

While the effectiveness of the combined high potential and rotation is self-evident from the results in **Fig. 4B**, it is understood that introducing such flows in more common battery configurations—*e.g.*, closed cells—would pose a potentially unsurmountable challenge. Alternative approaches exist for enhancing mass transport, which are compatible with constraints imposed by a closed battery cell. Specifically, it has been reported in previous studies that nonplanar electrode architectures may serve to reduce the local current density and thereby increase the effective mass transport at an electrode. Our previous work[7] also shows that, at a low current densities, nonplanar substrates (*e.g.*, interwoven carbon fibers) has negligible influence on the deposition morphology—the Li metal consistently grows into the moss-like structures, and does not form conformal deposition layer on the individual carbon fibers. **Fig. 4C~E** show the analogous results when Li is deposited on interwoven carbon fibers from 1M $LiPF_6$ in carbonate-based electrolyte. The deposition is performed at a large overpotential of -1.6 V versus $Li^+/Li$. The Li is seen to form a highly conformal deposition layer covering the surface of the individual carbon fibers. **Fig S9** shows a similar Li deposition morphology in an ether-based electrolyte. The consistency suggests this is a generalizable approach for eliminating the moss-like growth mode of Li. This morphology should be compared to Li deposition on carbon fibers at a lower current (see ref.[7]). This result, again, corroborates our finding that the moss-like growth pattern is due to the competing parasitic Faradaic reactions at low currents/potentials, and can be suppressed via rational designs of the electrokinetics. We further test the relevance of these discoveries to developing high-performance Li metal batteries. Stable Li plating/stripping behaviors over thousands of cycles are observed at a high current density of 70 mA/cm$^2$ (**Fig. 5A-5B**). Full-cell test using $LiFePO_4$ cathode illustrates a capacity retention of 80% over 350 cycles, with a charging current density of 70 mA/cm$^2$ and a discharging current density of 7 mA/cm$^2$ (**Fig. 5C-5D**).

**Discussion**

The results reported in the present study offer critical insights into the origin of moss-like electrodeposit growth during recharge of metal battery anodes. The first two of the three hypotheses normally used to explain non-planar metal deposition, *i.e.*, the surface diffusivity and the electrolyte cation diffusivity hypotheses—have been examined using a combination of a rotating disk electrode and electrodeposition at variable overpotentials. Our results show that neither are important in the regime where mossy deposition of metals is observed. Specifically, a smooth, compact metal electrodeposit morphology is expected at a low current density, where the surface diffusion and the electrolyte cation diffusion are relatively fast, and kinetical instabilities are suppressed. On the other hand, porous deposition morphology (*e.g.*, moss-like) should emerge at a high current density, where these two types of diffusion are slow relative to the deposition rate, and kinetical instabilities are therefore activated. The key findings from this study reported in **Figs. 2-4** contradict these expectations.

Our results are consistent with a third hypothesis—the chemical instability of the electrolyte. Specifically, upon electrodeposition, two competing reactions, *i.e.*, the reduction of the metal cations and the parasitic reduction of electrolyte components, occur simultaneously on the electrode surface. The solid products of these parasitic reactions, *e.g.*, polymers, inorganic compounds, etc., passivate the electrode surface non-uniformly, promoting "hot spots" in the growth landscape where the passivation layer is thinner and/or of a chemistry that facilitates deposition. As such, in considering the hypothesis, an additional degree of freedom is involved: electrolyte chemistry. The rate of parasitic reaction is directly dependent on the chemistry of the liquid electrolyte in contact with the electrode surface, and is therefore highly tunable by varying the electrolyte composition. This is consistent with our experimental observations that moss-like growth is predominant in electrolytes associated with a lower chemical stability and at small current densities where the rate of SEI-forming parasitic reactions is comparable to or higher than the rate of metal electrodeposition. The moss-like growth of metals in electroplating is therefore attributable to the competing Faradaic reactions—*i.e.*, the SEI-forming parasitic reactions and the metal electrodeposition reaction—occurring on the electrode surface.

**Conclusion**

In conclusion, comparing the electrokinetics-morphology relations manifested in the electrodeposition of Zn and Li in a group of representative electrolytes, we demonstrate unambiguously that the moss-like growth electrodeposit predominates in strongly-SEI forming electrolytes, particularly in the low-current density regime. In this regime, the competition between the Faradaic reactions—*i.e.*, SEI formation and metal deposition—gives rise to the non-uniform surface passivation and subsequently preferential proliferation of metal along high-flux paths.[24] Surprisingly, the persistent moss-like growth mode is readily suppressed at higher current densities, where the rate at which metal is deposited substantially exceeds—*i.e.*, is roughly one order of magnitude higher than the rate at which SEI-forming parasitic reactions occur. Consistent with results from proof-of-concept full cell battery tests reported, knowledge from the study uncovers—counterintuitively—the promise held by fast charged battery electrodes.

ACKNOWLEDGEMENTS

The authors express their gratitude to Dr. R. Luo for valuable discussions. **Funding**: This work was supported by DOE BES under award # DE-SC0016082 and as part of the Center for Mesoscale Transport Properties, an Energy Frontier Research Center supported by the U.S. Department of Energy, Office of Science, Basic Energy Sciences, under award #DE-SC0012673. This work made use of the Cornell Center for Materials Research Shared Facilities which are supported through the NSF MRSEC program (DMR-1719875). **Author Contribution**: L.A.A. directed the project. J.Z. and L.A.A. conceived and designed this work. J.Z., J.Y. and T.T. performed the electrodeposition, electrochemical measurements and structure characterizations. All the authors analyzed and discussed the data. J.Z. and L.A.A wrote the manuscript with important inputs from all the authors. **Data Availability**: All data needed to evaluate the conclusions in the paper are present in the paper and/or the Supplementary Materials. **Competing Interest Statement**: The authors declare no other competing interests.